\documentclass[aps,prl,epsfig,twocolumn]{revtex4}

\usepackage[dvips]{graphics}
\usepackage{graphicx}
\usepackage{amsfonts}
\usepackage{amssymb}
\usepackage{amsmath}

\begin{document}

\title{Stabilizing the Discrete Vortex of Topological Charge $S=2$}
\author{ P.G. Kevrekidis$^1$ and  
D.J.\ Frantzeskakis$^{2}$} 
\affiliation{ 
$^1$ Department of Mathematics and
Statistics, University of Massachusetts,
Amherst MA 01003-4515, USA \\
$^{2}$ Department of Physics, University of Athens, Panepistimiopolis,
Zografos, Athens 15784, Greece}

\begin{abstract}
We study the instability of the discrete
vortex with topological charge $S=2$ in a prototypical lattice model 
and observe its mediation through
the central lattice site. Motivated by this finding, we analyze 
the  model with the central site
being {\it inert}. We identify analytically and observe numerically the 
existence of a range of 
linearly stable discrete vortices with $S=2$ in the latter model. The range
of stability is comparable to that of the recently observed experimentally
$S=1$ discrete vortex, suggesting the potential for observation of
such higher charge discrete vortices.
\end{abstract}

\maketitle

{\it Introduction}. 
In the past decade, 
lattice systems described by differential-difference equations in which 
the evolution 
variable is continuum and the spatial variables are discrete, have been a subject of increasing interest 
\cite{reviews}.
These systems appear in many diverse physical contexts, describing, e.g.,      
the spatial dynamics of optical beams in coupled waveguide arrays in 
nonlinear optics \cite{reviews1}, 
the temporal evolution of Bose-Einstein condensates (BECs) 
in optical lattices 
in soft-condensed matter physics \cite{reviews2}, 
the DNA double strand in biophysics \cite{reviews3}, and so on. 

One of the principal directions of interest in these lattice systems
consists of the effort to understand the features of their
localized, solitary wave solutions. In two dimensions, such 
structures can be discrete solitons \cite{solit} or discrete
vortices (i.e., structures that have topological
charge over a discrete contour) \cite{vort}. In the past two years, there 
has been a considerable effort 
towards the observation 
of both entities in the context of optics, utilizing photorefractive crystals: regular discrete
solitons, dipole solitons, soliton-trains, soliton-necklaces 
and vector solitons were observed \cite{esolit}, while  
two groups were independently able to experimentally
produce robust discrete vortex states \cite{evort}. 
On the other hand, experimental developments in the physics of  
BECs closely follow with prominent recent results, including the observation
of bright, dark and gap solitons in quasi-one-dimensional settings
\cite{bec}, and with the generation of similar structures in higher dimensions
appearing within experimental reach \cite{ol2d}.

The above discussed experimental realization of discrete vortices
of topological charge $S=1$ 
\cite{evort} (i.e., with a 2$\pi$ phase shift around
a discrete contour) poses the question of whether higher topological
charge discrete vortices could also potentially be experimentally realizable.
The most natural higher topological charge state to consider is then
the vortex with $S=2$. However, lattice computations with a 
prototypical discrete model,
namely the the discrete nonlinear Schr{\"o}dinger (DNLS) equation, 
had identified that mode to be {\it always} unstable \cite{vort}. 
In this work we will revisit this topic and examine in some detail the instability of the
$S=2$ discrete vortex in the framework of the DNLS equation,
which, in different variants, is relevant 
to all of the above mentioned, experimentally tractable settings. 
Our scope is to offer some insight on the 
nature of the instability, which will, in turn, allow
us to suggest an explicit mechanism for its stabilization of this vortex by means of the inclusion of an impurity at
its center. We analyze the latter case in detail and establish (analytically and numerically)  
the stability of the $S=2$ vortex in that setting for parametric regimes similar to the ones 
for which the discrete vortex of $S=1$ has been found to be stable. 

{\it Analytical Results}. We consider 
the DNLS equation, 
\begin{eqnarray}
i \dot{u}_{n,m}= - \epsilon \Delta_2 u_{n,m} - |u_{n,m}|^2 u_{n,m},
\label{deq1}
\end{eqnarray}
where $u$ is the complex field (the envelope of the electric
field in optics or the wavefunction in BECs), $\epsilon$ is
the coupling constant (the ``tunneling rate'' between sites/wells),
while $\Delta_2 u_{n,m}=u_{n+1,m} + u_{n-1,m} + u_{n,m+1} + u_{n,m-1} - 4
u_{n,m}$ is the discrete Laplacian. We seek standing wave, localized
solutions in the standard form $u_{n,m}=\phi_{n,m} e^{i (\mu-4 \epsilon)
t}$. Our 
approach is based on the
Lyapunov-Schmidt theory for the existence of solutions \cite{golub} and
on linear stability analysis for tracing the stability eigenvalues
of the corresponding solutions, similarly to what was done for
discrete solitons and vortices in \cite{FKP}. 

Our starting point is the so-called anti-continuum limit of 
$\epsilon=0$ \cite{MA94}, where the nonlinear oscillators of our
two-dimensional lattice are uncoupled. We excite a
discrete vortex in that limit by choosing a contour containing
$8$ sites ((-1,-1), (-1,0), (-1,1), (0,1), (1,1), (1,0), (1,-1), (0,-1)),
where the corresponding solution is $e^{i \theta_{n,m}}$ on
these sites, while it is $0$ elsewhere 
(selecting without loss of generality $\mu=1$). 
The motivation for the choice of the phases over this discrete 
contour is that we 
aim to construct
a solution with $S=2$ over the relevant contour, hence the real
part of the configuration should behave as $\cos(2 \theta)$, while
the imaginary part as $\sin(2 \theta)$, which in turn immediately
implies that we should choose $\theta_{n,m}= j \pi/2$, where 
$j=1,\dots,8$ is an index over contour sites.
Below, we briefly discuss the general theory that would apply
to any solution over the relevant contour, and then focus on the
discrete vortex with $S=2$. The stationary state equation for
$\phi_{n,m}$ is given by:
\begin{eqnarray}
f(\phi_{n,m},\bar{\phi}_{n,m}, \epsilon)= (1-|\phi_{n,m}|^2) \phi_{n,m}
- \epsilon (\Delta_2+4) \phi_{n,m},
\label{deq2}
\end{eqnarray}
and its complex conjugate $\bar{f}(\phi_{n,m},\bar{\phi}_{n,m}, \epsilon)= 0$.
The linearization operator for these two difference equations
reads:
\begin{eqnarray}
\label{energy} {\cal H}_{n,m} &=& \left( \begin{array}{cc} 1 - 2
|\phi_{n,m}|^2 & - \phi_{n,m}^2 \\ - \bar{\phi}_{n,m}^2 & 1 - 2
|\phi_{n,m}|^2 \end{array} \right) 
\nonumber
\\
&-& \epsilon \left( s_{+1,0} +
s_{-1,0} + s_{0,+1} + s_{0,-1} \right) \left(
\begin{array}{cc} 1 & 0 \\ 0 & 1 \end{array} \right),
\end{eqnarray}
with $s_{n',m'}u_{n,m}=u_{n+n',m+m'}$.
Then, the solvability condition of the Lyapunov-Schmidt theory 
(allowing to continue a solution valid for $\epsilon=0$ to 
$\epsilon \neq 0$) mandates that the projection of the eigenvectors
of ${\cal H}_{n,m}^{\epsilon=0}$ to the Eq. (\ref{deq2}) and its
conjugate is null. To O$(\epsilon)$, this condition provides
the bifurcation function constraint 
$g_j^1=\sin(\theta_j-\theta_{j+1})+\sin(\theta_{j}-\theta_{j-1})=0$,
(for $j=1,\dots,8$ with periodic boundary conditions)
that was algebraically obtained in \cite{andrey}. This is 
naturally satisfied for 
the $S=2$ vortex with $\theta_j - \theta_{j-1}=
\pi/2$ discussed above. However, computing the Jacobian
matrix $({\cal M}_1)_{j,k} \equiv \partial g_j^1/\partial \theta_k$ 
of the bifurcation function $g^1$, 
one can 
observe that
its eigenvalues are 0 for the case of our $S=2$ solution and, hence, 
second-order reductions are necessary to 
adjudicate on
the existence/stability of the $S=2$ vortex. 

Expanding $\phi_{n,m}= \phi_{n,m}^0 + \epsilon \phi_{n,m}^1 + O(\epsilon^2)$,
one can obtain the corresponding equations for the O$(\epsilon)$
correction to the solution profile $\phi_{n,m}^0$ as:
\begin{equation}
\label{deq4} 
(1 - 2 |\phi_{n,m}^{0}|^2) \phi_{n,m}^{1} -
(\phi_{n,m}^{0})^{2} \bar{\phi}_{n,m}^{1} = (\Delta_2+4) \phi_{n,m}^0.
\end{equation}
The solution of Eq. (\ref{deq4}) can be found as 
\begin{eqnarray}
\label{deq5} \phi_{n,m}^{1} = - \frac{1}{2} \left[ \cos
(\theta_{j-1}-\theta_j) + \cos(\theta_{j+1}-\theta_j) \right] e^{i
\theta_j},
\end{eqnarray}
over the discrete contour  while $\phi_{0,0}^1=e^{i \theta_2} + e^{i
\theta_4} + e^{i \theta_6} + e^{i \theta_8}$. Using $\phi^1$ to
obtain the next order correction of the bifurcation function, we
get: $g_j^2=
\frac{1}{2} \sin (\theta_{j+1} - \theta_{j} ) \left[ \cos(\theta_j -
\theta_{j+1}) + \cos(\theta_{j+2} -\theta_{j+1}) \right] 
\label{sin-bifurcation2} +  \frac{1}{2} \sin(\theta_{j-1} -
\theta_{j}) \left[ \cos(\theta_j - \theta_{j-1}) + \cos(\theta_{j-2}
-\theta_{j-1}) \right]
+ \sin(\theta_j - \theta_{j+2}) + \sin(\theta_j -
\theta_{j+4}) (\delta_{j,2}+\delta_{j,4}+\delta_{j,6} 
+ \delta_{j,8})$, with $1 \leq j \leq 8$ and $\delta$ denoting 
the Kronecker symbol.  Once again the bifurcation condition
is satisfied for our vortex of $S=2$. However, the eigenvalues
of the corresponding second-order Jacobian ${\cal M}_2$
are not identically zero and can be used to establish (in conjunction
with the bifurcation condition being identically satisfied)
the persistence of the vortex of topological charge $S=2$ in the
vicinity of $\epsilon=0$. 

Furthermore, the Jacobian ${\cal M}_2$ of the second order reductions
can be computed explicitly as:
\begin{equation}
{\cal M}_2 = \left( \begin{array}{ccccccccc}
1 & 0 & -\frac{1}{2} & 0 & 0 & 0 & -\frac{1}{2} & 0 \\
0 & 0 & 0 & \frac{1}{2} & 0 & -1 & 0 & \frac{1}{2}  \\
-\frac{1}{2} & 0 & 1 & 0 & -\frac{1}{2} & 0 & 0 & 0 \\
0 & \frac{1}{2} & 0 & 0 & 0 & \frac{1}{2} & 0 & -1 \\
0 & 0 & -\frac{1}{2} & 0 & 1 & 0 & -\frac{1}{2} & 0 \\
0 & -1 & 0 & \frac{1}{2} & 0 & 0 & 0 & \frac{1}{2} \\
-\frac{1}{2} & 0 & 0 & 0 & -\frac{1}{2} & 0 & 1 & 0 \\
0 & \frac{1}{2} & 0 & -1 & 0 & \frac{1}{2} & 0 & 0
\end{array} \right).
\end{equation}
By using the expansion 
\begin{equation}
\label{linearization} u_{n,m}(t) = e^{i (1 - 4 \epsilon) t + i
\theta_0} \left( \phi_{n,m} + a_{n,m} e^{\lambda t} + \bar{b}_{n,m}
e^{\bar{\lambda} t} \right),
\end{equation}
one can study the stability of the discrete vortex of $S=2$. 
Furthermore by expanding the eigenfunction in Taylor
series in $\epsilon$ (as we did above for the solution $\phi$)
and correspondingly the eigenvalue $\lambda$ as $\lambda=
\epsilon \lambda_1 + O(\epsilon^2)$, it can be shown (see
\cite{FKP} for details) that the Jacobian ${\cal M}_2$ can
be directly connected with the eigenvalue correction 
$\lambda_1$ (for eigenvalues bifurcating from 0, which are
the natural sources of potential instability in the DNLS problem).
The relevant equation connecting ${\cal M}_2$ and $\lambda_1$
is the (reduced, i.e., $8 \times 8$) eigenvalue problem of the form:
\begin{eqnarray}
{\cal M}_2 {\bf c} = \lambda_1
{\cal L}_2 {\bf c} + \frac{\lambda_1^2}{2} {\bf c},
\label{deq6}
\end{eqnarray}
where $({\cal L}_2)_{j,k}=1$ if $j=k-1$, $-1$ if $j=k+1$ and
$0$ if $|j-k| \neq 1$.
Using the discrete Fourier transform, one can
obtain from (\ref{deq6}) the
characteristic polynomial:
\begin{eqnarray}
\nonumber
\lambda_1^4 &-& 2 \lambda_1^2 \left( 1 - (-1)^j - 8 \sin^2 \frac{\pi
j}{4} \right) 
+ 8 \sin^2 \frac{\pi j}{4}  \times
\\
( 1 &-& (-1)^j - 2
\sin^2 \frac{\pi j}{4}) = 0, \qquad j = 1,2,3,4,
\label{deq6a}
\end{eqnarray}
which provides the leading order approximations to the eigenvalues
of the $S=2$ vortex as follows: in the neighborhood of 
$\lambda=0$, the vortex will have three eigenvalues of algebraic
multiplicity four: $\lambda=0$ and $\lambda=\pm \sqrt{2} \epsilon i$,
while it will have two simple imaginary eigenvalues
$\lambda=\pm \sqrt{\sqrt{80}+8} \epsilon i$ and two real
eigenvalues $\lambda= \pm \sqrt{\sqrt{80}-8} \epsilon$. 
Among the latter, the positive one is the reason for the
$S=2$ vortex being {\it always} (i.e., for any
$\epsilon \neq 0$) unstable, as was numerically observed
\cite{vort}.

The examination of the real eigenmode leading to the
direct instability of the $S=2$ vortex (that has support
over the central, i.e., (0,0), site), as well as the apparent
mediation (in numerical experiments--see below) of the 
instability by means of the central site, lead us to consider
the possibility of having an ``impurity'' at the central
site, e.g., a strong
localized potential such as a laser beam in BECs or an 
inhomogeneity in the photorefractive crystal, enforcing $\phi_{0,0}=0$.  
In such a case, 
the bifurcation function $g_j^2$ lacks the last term
(encompassing the Kronecker symbols), since these are 
interactions ``mediated'' by the now inert (0,0) site. 
Furthermore, the second order Jacobian is now much simpler
and acquires the form: $({\cal M}_2)_{j,k}=1$ for $j=k$,
$-1/2$ for $j=k \pm 2$ and $0$ for $|j - k | \neq 0,2$.
One can then repeat the calculation of the eigenvalues in
the problem of Eq. (\ref{deq6}), via the discrete Fourier
transform, to obtain 
the characteristic equation:
\begin{eqnarray}
\left(\lambda_1 + 2 i \sin(\frac{j \pi}{4})\right)^2=0, \qquad j=1,\dots,8.
\label{deq7}
\end{eqnarray}
This results into three eigenvalues of algebraic multiplicity
four, namely $\lambda=0$ and $\lambda= \pm \epsilon i \sqrt{2}/2$.
There are also two double eigenvalues $\lambda= \pm 2 i$. The
crucial observation, however, is that in this case, there are
no real eigenvalues immediately present as $\epsilon \neq 0$ 
and hence the discrete vortex with $S=2$ will be {\it linearly
stable}, due to the stabilizing effect of the
impurity (or, to be more precise, due to the absence of the
instability mediated by the (0,0) site). We now turn to numerical
investigations to examine the validity of these findings.

{\it Numerical Results}. We identify unit frequency solutions with topological
charge $S=2$, by initializing the exact solution configuration
at the $\epsilon=0$ limit of Eq. (\ref{deq1}) and then using continuation over 
$\epsilon$, combined with a contraction mapping for the solution
of the nonlinear system of (algebraic) equations
to identify the
exact (up to a prescribed numerical accuracy) numerical discrete
vortex. We then perform linear stability analysis, using the
expansion of Eq. (\ref{linearization}), to obtain the eigenvalues
$\lambda$, and their corresponding eigenvectors.

\begin{figure}[tbp]
\begin{center}
\begin{tabular}{lll}
\hskip-0.2cm\includegraphics[width=2.95cm]{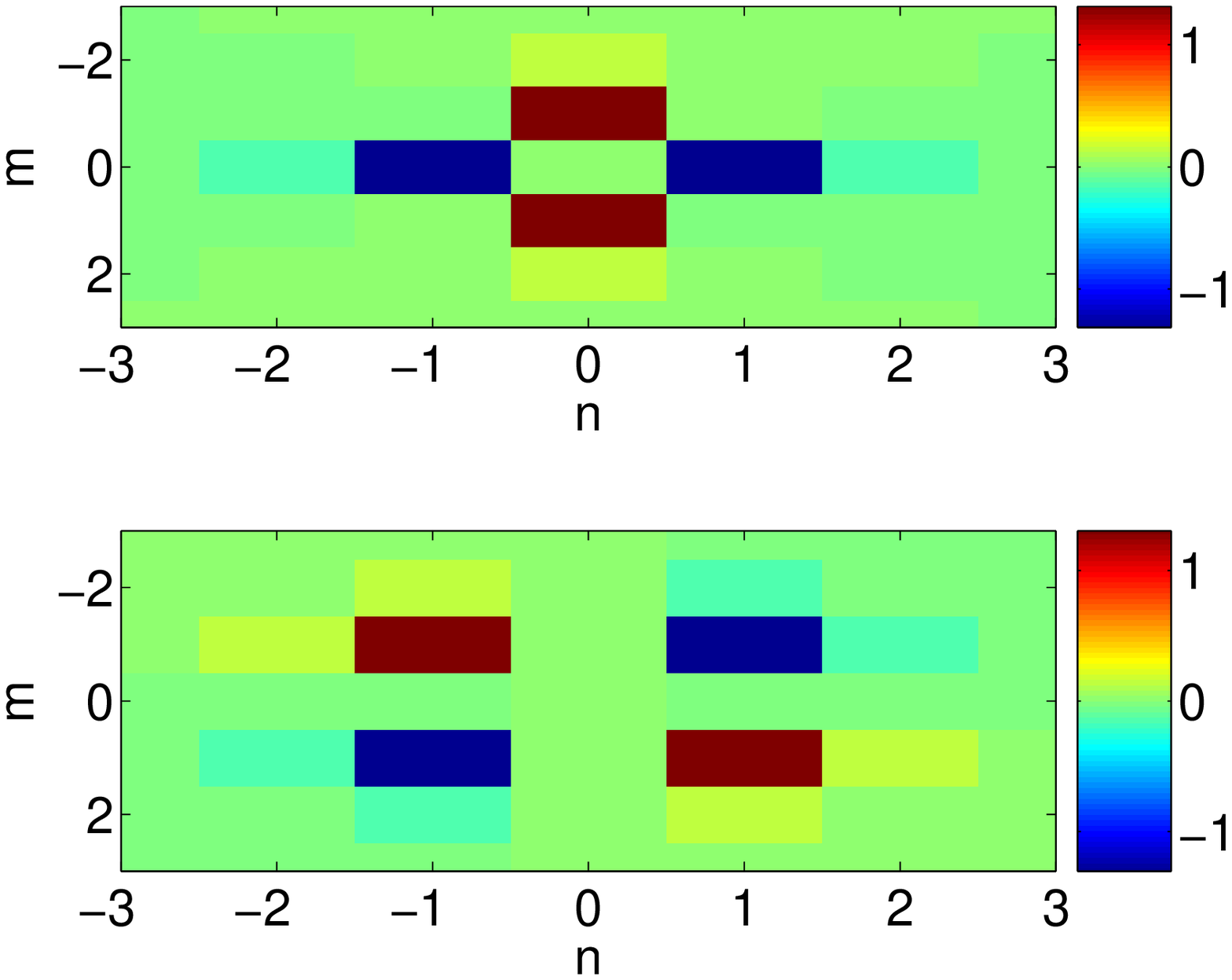} &
\hskip-0.2cm\includegraphics[width=2.95cm]{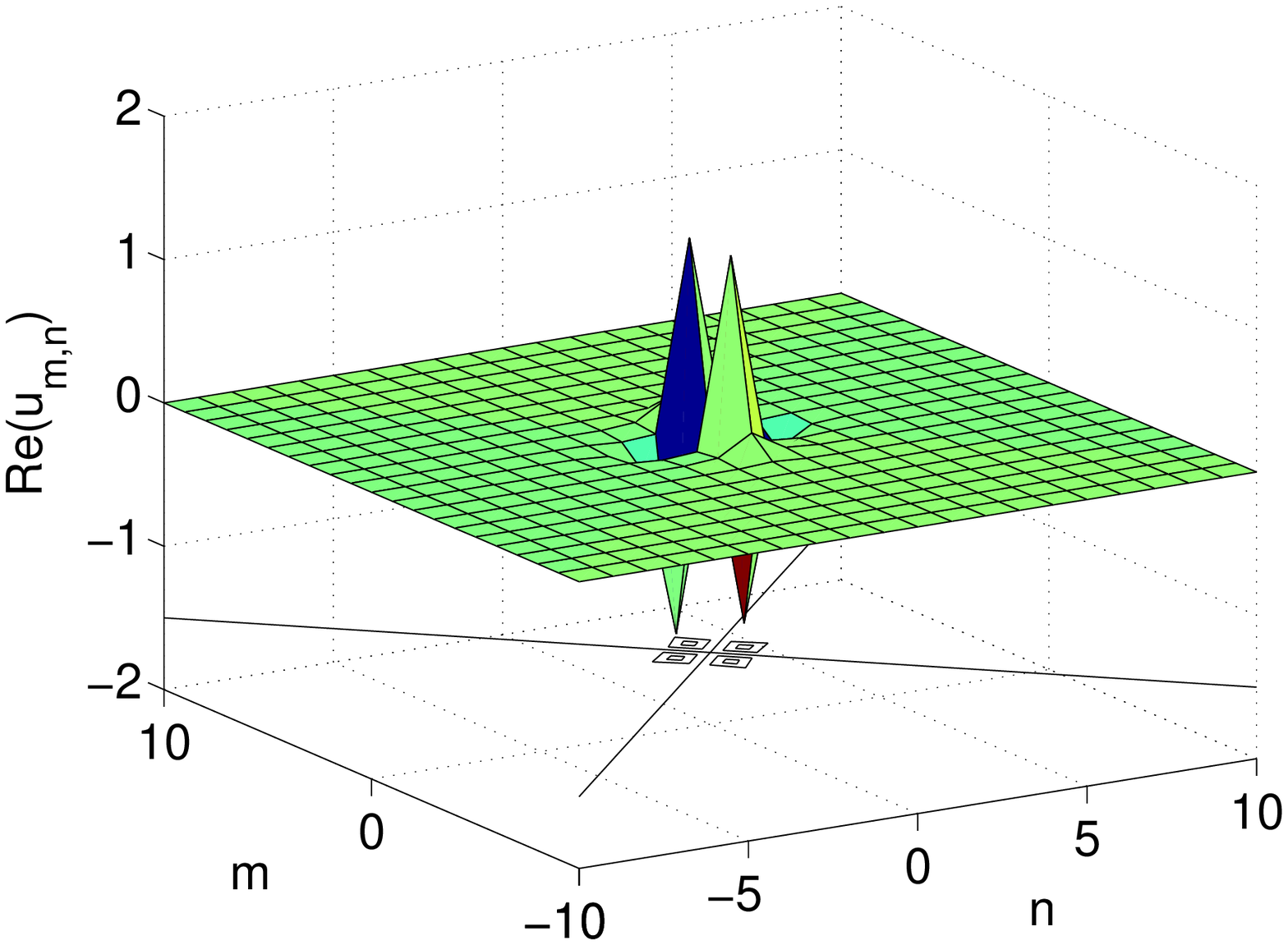} &
\hskip-0.2cm\includegraphics[width=2.95cm]{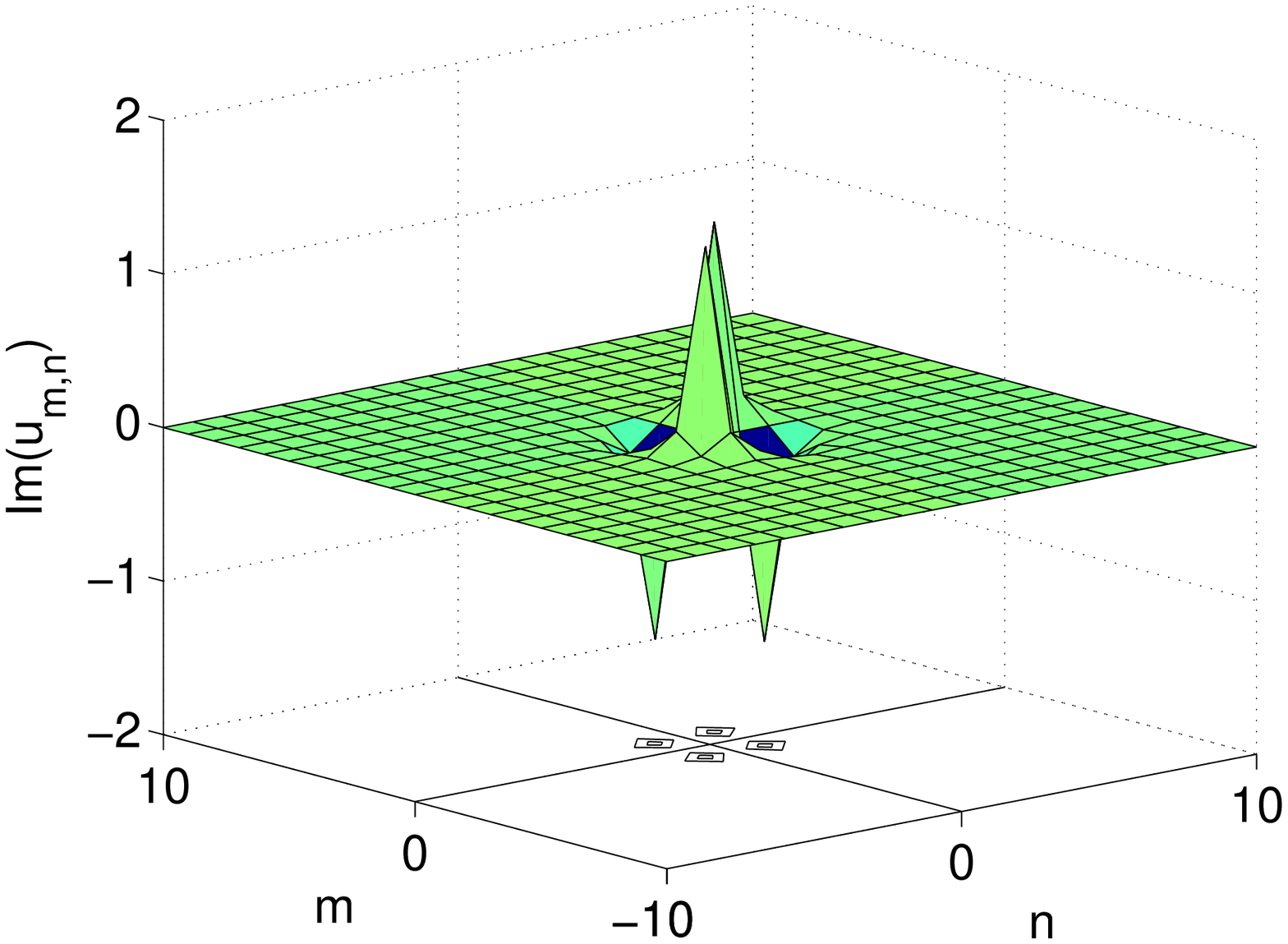} \\[-0.0ex]
\end{tabular}
\end{center}
\caption{The discrete vortex is shown for $\epsilon=0.2$. The left
panel shows contour plots of the real (top) and imaginary (bottom)
parts. The middle panel shows a 3d rendering of the real part,
while the right panel shows a similar 3d plot of the imaginary part
of the vortex.}
\label{Fig1}
\end{figure}

\begin{figure}[tbp]
\begin{center}
\begin{tabular}{ll}
\hskip-0.2cm\includegraphics[width=4.2cm]{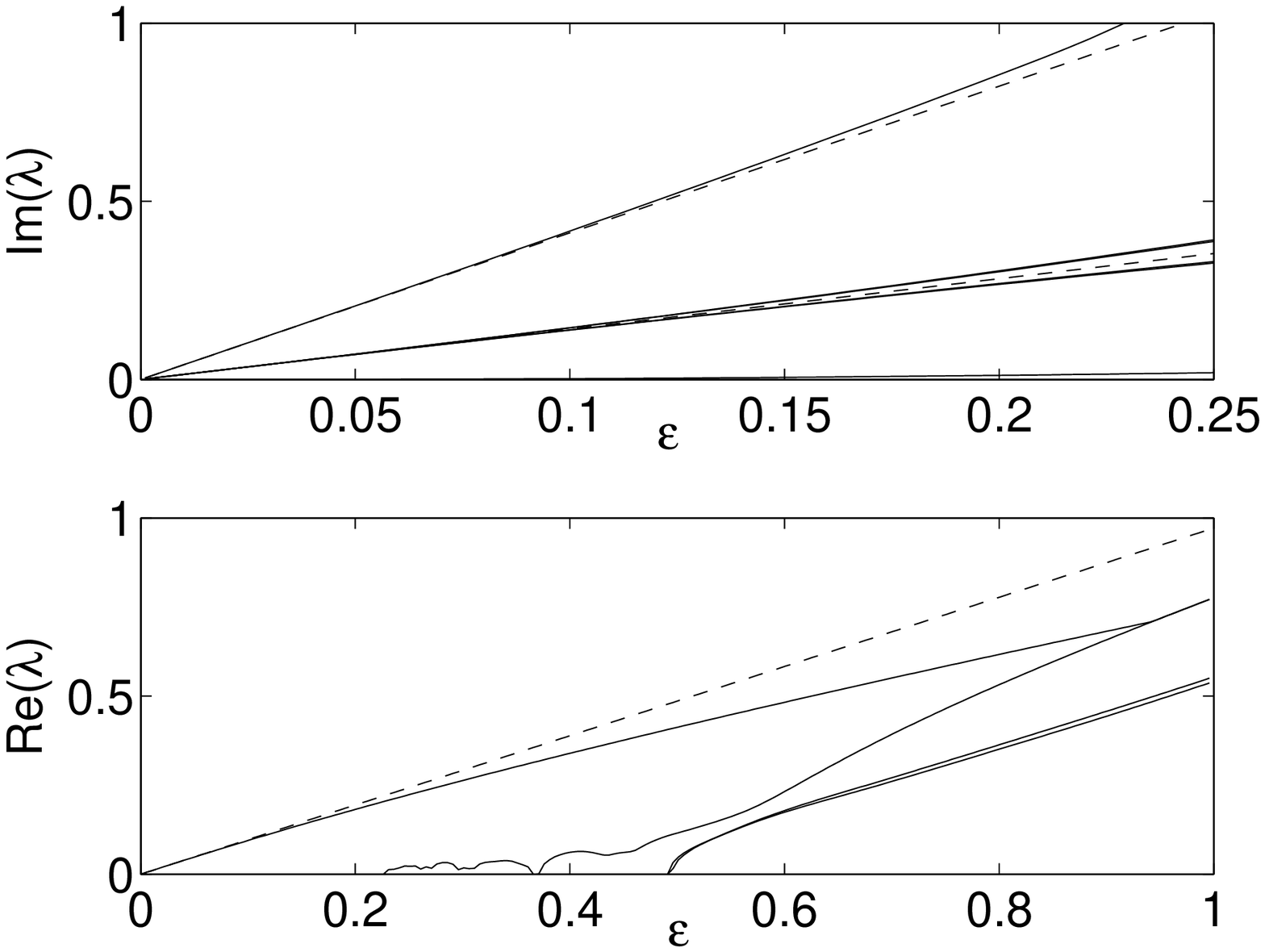} &
\hskip0.0cm\includegraphics[width=4.2cm]{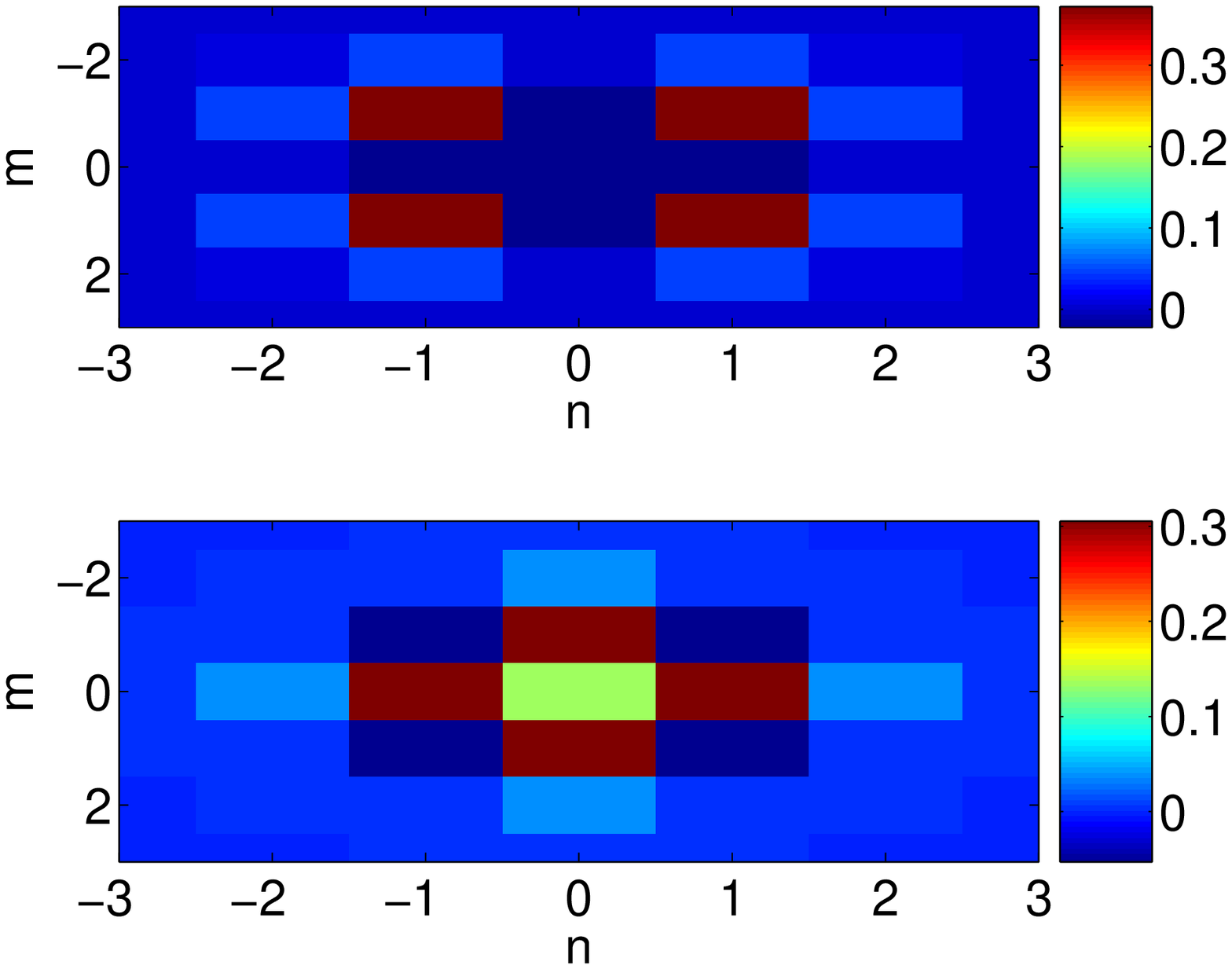} \\[-5.0ex]
\end{tabular}
\end{center}
\caption{Linear stability analysis of the $S=2$ vortex. The left
panel shows the imaginary (top) and the real (bottom) part of
the point spectrum eigenvalues bifurcating from $\lambda=0$. 
The solid line shows the numerical result, while the dashed line
indicates the theoretical prediction (see text). The most unstable
eigenmode pertaining to the real eigenvalue for $\epsilon=0.2$
is shown (real part: top; imaginary part: bottom) in the right panel.}
\label{Fig2}
\end{figure}

Figure \ref{Fig1} shows a typical example of the discrete vortex
for $\epsilon=0.2$ in the regular DNLS model. The middle and
right panels show the real and imaginary part of the solution,
clearly emulating $\cos(2 \theta)$ and $\sin(2 \theta)$ over
the lattice contour of interest. The linear stability analysis
of this vortex is shown in Fig. \ref{Fig2}. One can observe that
both for the imaginary  (top left panel), as well as for the real
(bottom left panel) eigenvalues, the predictions of the perturbation theory
(dashed line) are extremely accurate in comparison with the
full numerical results even for $\epsilon$ up to $0.25$. Clearly
as $\epsilon$ increases, higher order phenomena become relevant
such as the splitting of the quartet of eigenvalues at
$\lambda=\pm \sqrt{2} \epsilon i$, or the collision of the
simple pair of eigenvalues with $\lambda=\pm \sqrt{\sqrt{80}+8} \epsilon i$
with the bottom edge of the continuous spectrum (which is at
$\lambda = \pm i$), resulting into a Hamiltonian Hopf bifurcation
and a complex quartet of eigenvalues for $\epsilon > 0.23$. Additional
such quartets appear for larger values of $\epsilon$. However,
the solution is always unstable due the real eigenvalue pair
$\lambda= \pm \sqrt{\sqrt{80}-8} \epsilon$, whose eigenfunction
is shown in the right panel of the figure. Notice that the latter
has support over the central site, predisposing us for the role
of this site in the instability development.


\begin{figure}[tbp]
\begin{center}
\begin{tabular}{ll}
\hskip-0.2cm\includegraphics[width=4.2cm]{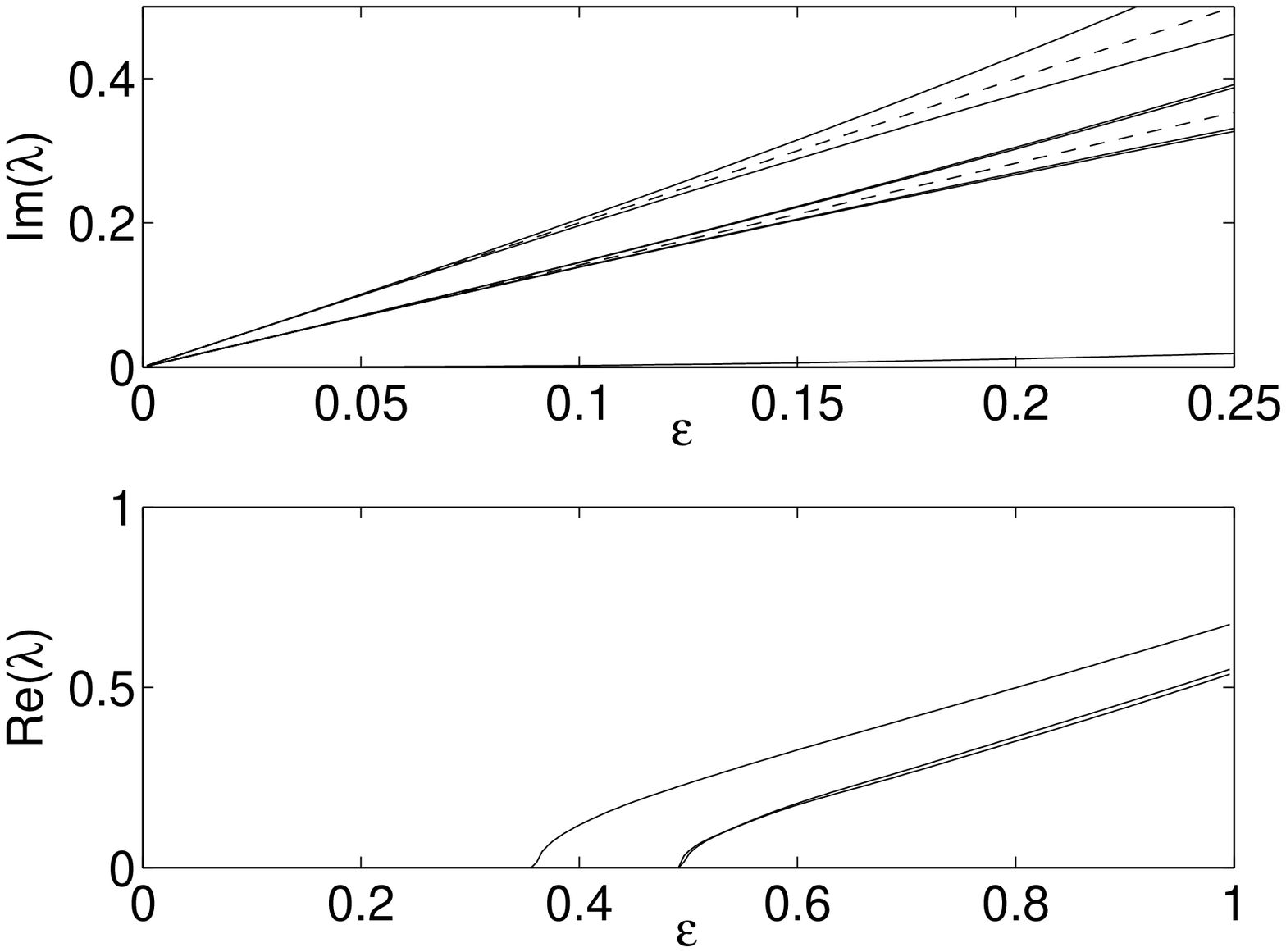} &
\hskip0.0cm\includegraphics[width=4.2cm]{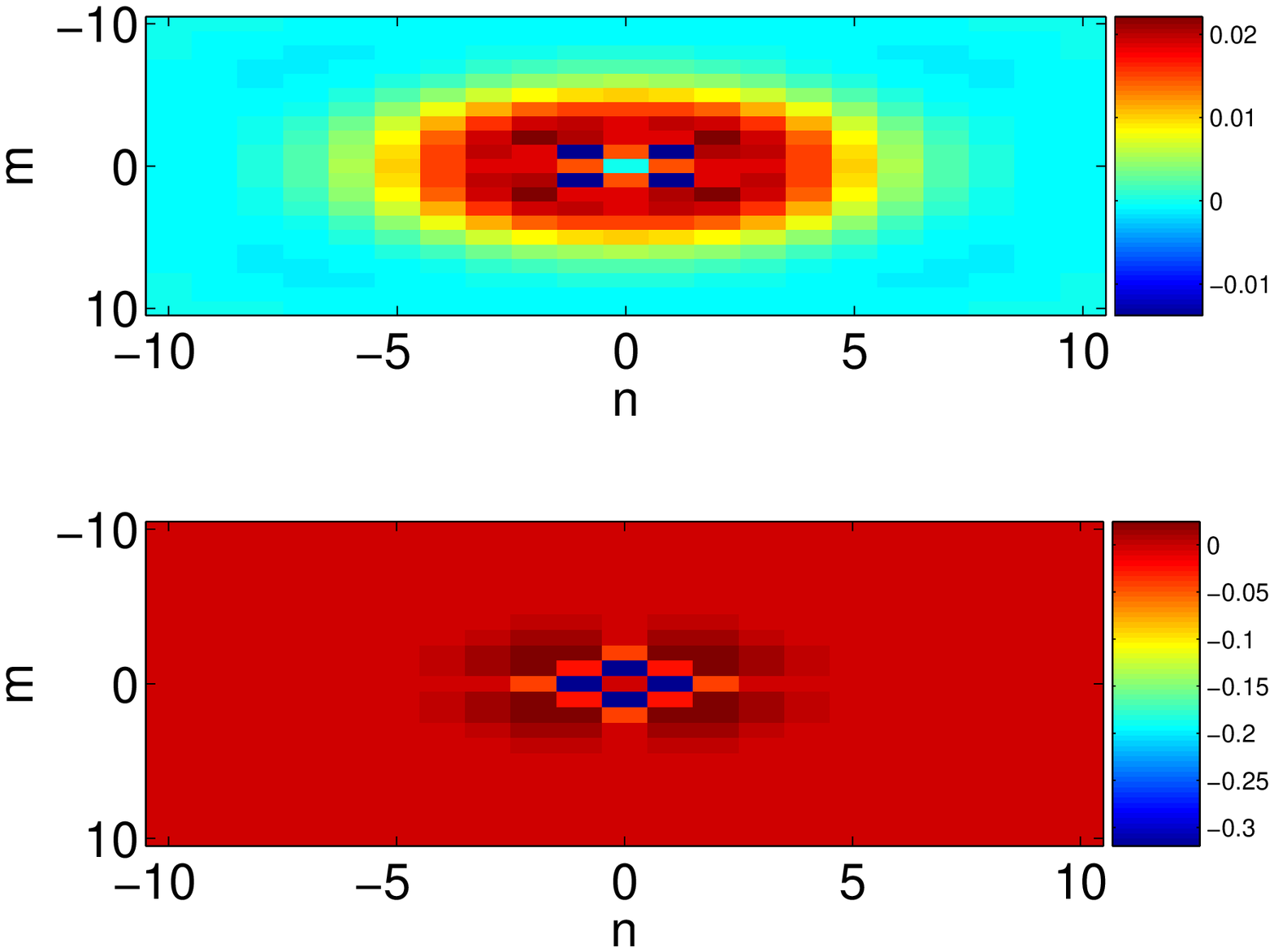} \\[-5.0ex]
\end{tabular}
\end{center}
\caption{Same as in Fig. 2, but for the model with the 
impurity (i.e., with (0,0) inert). The instability is absent
in this case. The right panel shows
the principal eigenmode of instability for $\epsilon=0.4$.}
\label{Fig3}
\end{figure}

The corresponding predictions/numerical results for the model with the 
impurity (i.e., with (0,0) inert) are shown in Fig. \ref{Fig3}.
The top panel illustrates the eigenvalue of multiplicity four
with $\lambda= \pm \epsilon i \sqrt{2}/2$ and with multiplicity
two $\lambda= \pm 2 i$, which are again in excellent agreement
with the numerical findings. The real part clearly indicates the
{\it absence} of an instability for small $\epsilon$. Such an
instability arises due to collision of the eigenvalue pair with the
continuous spectrum and is present for $\epsilon>0.36$. It
is crucial to note here that the $S=1$ vortex was found to
be stable for $\epsilon < 0.38$ in \cite{FKP}. This illustrates
that the present mechanism stabilizes the $S=2$ vortex for a
parametric region comparable to that of the $S=1$ vortex, hinting 
that it could be experimentally feasible to trace such a configuration
similarly to what was done for the $S=1$ case \cite{evort}.
The right panel shows the real and imaginary part of an unstable
eigenmode for $\epsilon=0.4$.


\begin{figure}[tbp]
\begin{center}
\begin{tabular}{ll}
\hskip-0.2cm\includegraphics[width=4.3cm]{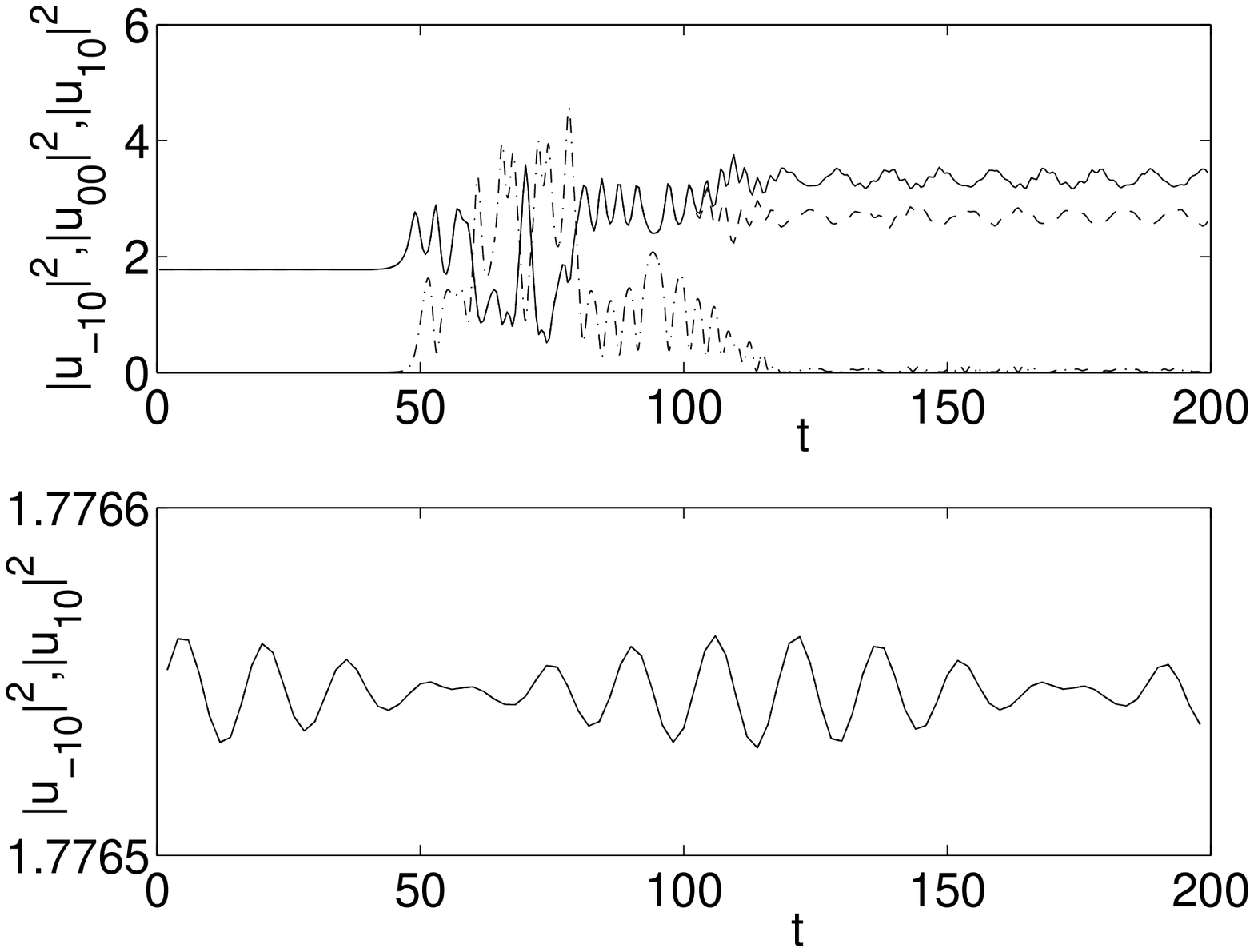} &
\hskip0.0cm\includegraphics[width=4.3cm]{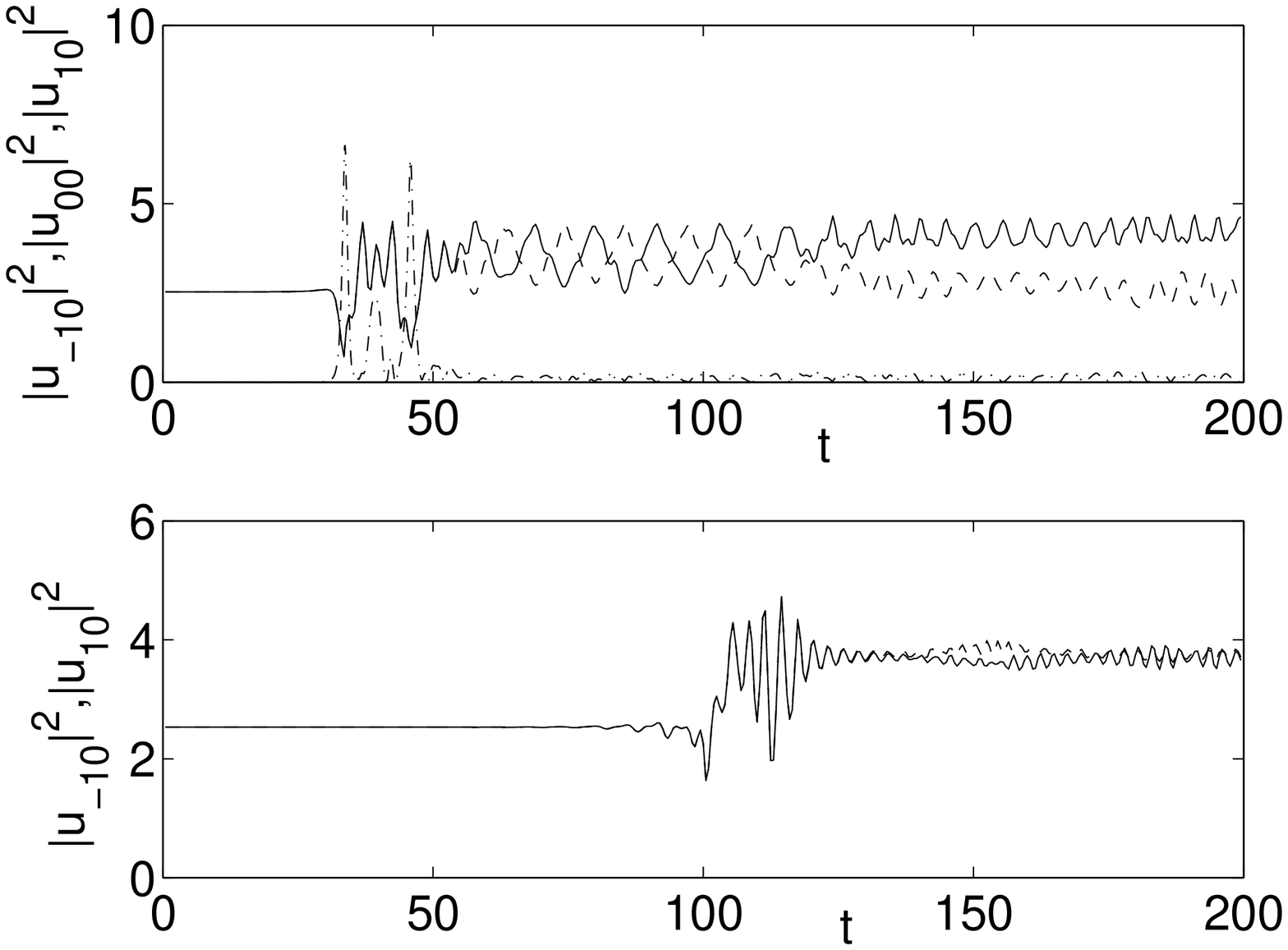} \\[-5.0ex]
\end{tabular}
\end{center}
\caption{For identical initial conditions (see text), 
 $\epsilon=0.2$ (left) and $\epsilon=0.4$ (right), 
the solution is shown in the relevant central sites
as a function of time, in the absence (top) and presence (bottom)
of the impurity. 
The central site (dash-dotted line) and the sites $(-1,0)$ (solid line)
and $(1,0)$ (dashed line) are shown (the latter two 
are the only ones that remain
excited in the configuration for long times).}
\label{Fig4}
\end{figure} 


Finally, to examine the dynamical development of the instability
and to compare/contrast the dynamical features of the two models
(in the absence and presence, respectively, of the impurity), we have
conducted direct numerical experiments. The main results are shown in Fig. \ref{Fig4} 
for two
representative cases (namely $\epsilon=0.2$, where the former 
case is unstable, while the latter is stable, and $\epsilon=0.4$,
where both models have unstable $S=2$ vortices). In both cases,
we have simulated both models up to $t=200$, initializing them
with {\it identical} initial conditions consisting
of the vortex with a perturbation (multiplied by $10^{-4}$) 
in the eigendirection of  the right panel of 
Fig. \ref{Fig2} for $\epsilon=0.2$ and of Fig. \ref{Fig3} for $\epsilon=0.4$. 
In the case of $\epsilon=0.2$, we observe that the DNLS vortex
becomes unstable, whereas in the presence of the impurity the
solution is completely stable (exhibiting oscillations at the order of the
initial perturbation). The instability for the DNLS vortex appears
to be mediated by the central site (dash-dotted) line, which eventually
settles at a rather small amplitude. The final configuration finds 
6 of the 8 (initially) 
participating sites at the vortex with near-zero amplitudes,
while only two sites (shown by solid and dashed line) remain excited
in an asymmetric configuration with a long-lived breathing (weak)
exchange of power between them, mediated principally by the central site.
The instability sets in for $t \approx 45$ in this case. For $\epsilon=0.4$,
the regular DNLS becomes unstable even faster ($t \approx 30$) and
the dynamics is similar. For the case with the impurity the instability
sets in at much longer times ($t \approx 80$), as expected by the
much smaller value of the corresponding principal eigenvalue real part.
Furthermore, while only two sites remain excited in this case as well,
the configuration is no longer asymmetric and the oscillation of
power can be identified (data not shown) as being caused by the  
small amplitude exchange of power around  the vortex contour 
(recall that in this case the central site is in this case inert). 

{\it Conclusions}. In this work we have revisited the topic of
discrete vortices of topological charge $S=2$. We have explicitly
discussed and illustrated their instability in the prototypical
lattice model of the discrete nonlinear Schr{\"o}dinger equation
and have traced its source in the exchange of power made available
through the central site of the vortex. We have thus proposed to
consider a model with an impurity (an inert) site at the center of
the vortex. Examination of the stability problem in the latter
case shows the absence of linear instability for a regime of
coupling strengths comparable to that of the linear stability
interval of the experimentally observable $S=1$ state. Numerical
findings conclusively corroborate this picture both at the level
of linear stability analysis (found to be in excellent agreement with the
theoretical predictions) and at the one of direct numerical 
experiments. We believe that this opens the path for observation
of higher charge discrete vortices and renders experimental work in this
direction particularly timely.

Numerous constructive discussions and comments by D.E. Pelinovsky
and partial support by  NSF-DMS-0204585, NSF-CAREER, and the 
Eppley Foundation for Research (PGK) are gratefully acknowledged.

\end{document}